\documentclass[%
 reprint,
superscriptaddress,
 amsmath,amssymb,
 aps,
 pre,
]{revtex4-1}

\usepackage{graphicx}
\usepackage{dcolumn}
\usepackage{bm}
\usepackage[colorlinks= true, linkcolor=blue, citecolor=blue, urlcolor=blue]{hyperref}
\usepackage{lipsum}
\usepackage{soul}
\usepackage{bbold}
\usepackage[normalem]{ulem}
\usepackage{mathtools}
\usepackage{xcolor}
\usepackage{mathtools}
\usepackage{amsmath}

\begin{document}
\newcommand{\red}{\textcolor{red}}
\newcommand{\blue}{\textcolor{blue}}

\title[sample title]{Depolarization of opinions on social networks through random nudges}
\author{Ritam Pal}
\email{ritam.pal@students.iiserpune.ac.in}
\author{Aanjaneya Kumar}
\email{kumar.aanjaneya@students.iiserpune.ac.in}
\author{M. S. Santhanam}
\email{santh@iiserpune.ac.in}
\affiliation{Department of Physics, Indian Institute of Science Education and Research, Dr. Homi Bhabha Road, Pune 411008, India.}

\begin{abstract}
Polarization of  opinions has been empirically noted in many online social network platforms. Traditional models of opinion dynamics, based on statistical physics principles, do not account for the emergence of polarization and echo chambers in online network platforms. A recently introduced opinion dynamics model that incorporates the \emph{homophily} factor -- the tendency of agents to connect with those holding similar opinions as their own -- captures polarization and echo chamber effects. In this work, we provide a non-intrusive framework for mildly nudging agents in an online community to form random connections. This is shown to lead to significant depolarization of opinions and decrease the echo chamber effects. Though a mild nudge effectively avoids polarization, overdoing this leads to another undesirable effect, namely, radicalization. Further, we obtain the optimal nudge probability to avoid the extremes of polarization and radicalization outcomes.
\end{abstract}

\maketitle


\section{Introduction}
The information revolution has lowered the entry barrier for nearly everyone to participate and contribute to shaping opinions and policies on various issues. This has been largely aided by the easy availability of social media infrastructure through mobile devices. Increasingly, the collective opinions expressed through various social media platforms are thought to be one barometer of the public mood on any contentious issue of the day \cite{social-media-as-public-opinion}. This provides an interesting testing ground for the dynamics and statistical physics of interacting multi-agent systems since the online nature of interactions provides fine-grained data for quantitative analysis and comparison with model results.

The study of opinion formation and its dynamics has attracted researchers for decades. The analysis of opinion dynamics from the statistical physics perspective can be traced back to the work of DeGroot \cite{reaching-a-consensus}, which provides a framework for reaching a consensus. Other discrete models, including the voter \cite{the-voter-model, reality-inspired-voter-models-a-mini-review} model, Sznajd model \cite{opinion-evolution-in-closed-community, sznajd-review}, and their variants which have a strong basis in a framework of interacting spins, suggest that large participatory interactions among agents might also lead to the emergence of consensus. However, empirical results have shown that the distribution of opinions tends to show a bimodal distribution pattern corresponding to polarization, especially on controversial issues of the day \cite{biased-assimilation-and-attitude-polarization, have-americans-social-attitudes-become-more-polarized, paritisans-without-constrait-political-polarization-and-trends}. Culture dissemination model \cite{the-dissemination-of-culture}, one of the first higher-dimensional modeling approaches to opinion dynamics, which also incorporates the human tendency to interact with similar persons, shows that despite there being local convergence, global polarization can be reached. Different variants of the bounded confidence model \cite{mixing-beliefs-among-interacting-agents, opinioin-dynamics-and-bounded-confidence} can also capture many empirically found trends in the distribution of opinions. These models can reproduce consensus, bimodal, or multi-modal opinion distributions depending on the confidence interval.

Another empirical feature that could not be accounted for by early models (at least by their original versions) was the phenomenon of echo chambers \cite{echo-chambers-online}. This refers to a scenario in which one agent's opinion is similar to the agents in their ``social neighborhood'', and one tends to reinforce the other. Lack of sufficient engagement with opposing opinions leads to positive reinforcement of one's own opinion within a close-knit social network. Empirical evidence for this effect has been reported from several social media platforms \cite{echo-chambers-emotional-contagion-and-group-polarization-on-facebook, quantifying-echo-chamber-effects-in-information-spreading-over-political-communication, political-discourse-on-social-media-echo-chambers, The_echo_chamber_effect_on_social_media}. Few recent opinion dynamics models \cite{modeling-echo-chambers-and-polarizaiton-dynamics-in-social-networks, polarized-idoelogy, link-recommendation-algorithms-and-dynamics-of-polarization-in-social-networks, social-influence-and-unfollowing-accelerate-the-emergence-of-echo-chambers} have qualitatively captured the features of echo chambers, which have been shown to arise from personalized interactions among peers in an online setting, which might be accelerated through the platform's recommendation engine. 

The model introduced by Baumann et al. accounted for several observed features from empirical data along with echo chambers in social media. The features that (a) most active users tend to be strongly opinionated and (b) locally connected agents have a convergence of opinions can be linked to the mechanism of reinforcement of opinion among agents and the tendency of agents to interact more with those with similar opinions (homophily \cite{birds-of-a-feather-homophily-in-social-networks, homophily-and-polarization-in-the-age-of-misinformation}). Even if the model starts from an initial distribution of opinions without clear preferences, highly homophilic interactions induce the formation of echo chambers and polarized states.

Though having diverse opinions might be a desired outcome, extreme polarization leads to network segregation \cite{segregatioin-and-clustering}, which often bottlenecks the information flow in social networks. Also, echo chambers, often linked to polarization, are known to be responsible for sustaining misinformation for a longer time on social networks \cite{echo-chambers-and-viral-misinformation, the-spreading-of-misinformation-online}. These problems call for intervention mechanisms, which should be safe and non-invasive. 

It might appear that in the case of controversial topics, the interaction and the debate will always lead to polarized states of opinion. But the underlying mechanism for polarization, the reinforcement of opinions through interaction between like-minded people, leaves us wondering if any intervention will help to reconcile disparate opinions.

In this work, we show that if agents are nudged slightly, then the cycle of reinforcement of opinions can be broken, and depolarization can be achieved. In social networks, the nudges are effected by exposing the agents to diverse opinions. We also show that overdoing this leads to radicalization \cite{the-group-polarization-phemomenon, group-polarization-a-critical-review-and-meta-analysis}, a state where all the agents have the same stance on an issue. We formulate an optimization problem that avoids polarization and radicalization and computes the right amount of nudge probability required to achieve this optimal scenario.

In the next section, we discuss the basic model and motivate the random nudges in the subsequent section. In Sec.IV, we demonstrate our results and discuss their implications. We formulate an optimization problem in Sec. V, which emerges from a trade-off between depolarization due to the proposed random nudges and the tendency to move towards a radicalized state. We conclude with a discussion of future directions.

\section{Basic Model and Methods}
To analyze polarization and to introduce possible intervention methods for reducing polarization, we adapt a recently introduced model for opinion dynamics \cite{modeling-echo-chambers-and-polarizaiton-dynamics-in-social-networks}. This model qualitatively captures a few aspects of opinion dynamics when agents' opinions evolve due to interactions in social media platforms. The model can reproduce the empirical features such as polarization and echo chambers and the fact that more active people on social media tend to have extreme opinions. 

The model has $N$ interacting agents, and it is assumed there are only two possible sides to an issue. This is typical of many, but not all, the issues -- for example, to allow abortion or not. Opinion on a given issue is denoted by $x_i$, which can take any real value in the range $(-\infty, \infty)$. The sign of the $x_i$ corresponds to the stance of the agent in the corresponding issue, and $|x_i|$ denotes the conviction of the agent in their respective stance. This implies that the larger the value of $|x_i|$, the more extreme the agent's opinion is. The model used to capture the evolution of opinion is activity driven \cite{activity-driven-modeling-of-time-varying-networks, topological-properties-of-time-integrated-activity-driven-netowork, burstiness-and-aging-in-social-temporal-networks, controlling-contagion-processes-in-activity-driven-networks}, {\it i.e.}, at each time step, only active agents are allowed to interact with other agents. Based on empirical data \cite{activity-driven-modeling-of-time-varying-networks, burstiness-and-aging-in-social-temporal-networks}, the probability for agents to be active is chosen to be 
\begin{equation}
    \label{activities.eq}
      F(a) = \frac{1-\gamma}{1-\varepsilon^{1-\gamma}} a^{-\gamma},
\end{equation}
where $a$ is the activity, $\varepsilon$ is the minimum activity (chosen in this work to be $10^{-2}$), and $\gamma$ controls how steep the function $F(a)$ which is chosen to be $2.1$. Agents' opinions evolve based on their interactions with other agents, and this information is encoded in the time-dependent adjacency matrix $A_{i, j}(t)$. Further, opinion evolution also depends on the strength of social interaction $K > 0$ and the controversialness of the issue $\alpha > 0$. The opinion dynamics is given by 
the following $N$ coupled differential equations \cite{modeling-echo-chambers-and-polarizaiton-dynamics-in-social-networks}
\begin{equation}
    \label{main.eq}
    \dot{x}_i= -x_i + K \left(\sum^{N}_{j=1} A_{ij} (t)  \tanh{(\alpha x_j)}\right).
\end{equation}
In this, $A_{i, j}(t)$ is the temporal adjacency matrix of interaction at time $t$. If at time $t$ agent $j$ influences agent $i$, than $A_{i, j}(t) = 1$, and $A_{i, j}(t) = 0$ otherwise. If agent $i$ is active at time $t$, they will interact with $m$ other agents, weighted by the probability $P_{i, j}$. Further, the probabilistic reciprocity factor $r \in [0, 1]$ determines the chance that an interaction is mutually influential, {\it i.e.}, $A_{ij}(t)=A_{ji}(t)=1$. The interaction probability is defined to be a function of the magnitude between two agents' opinions.
\begin{equation}
    \label{homophily.eq}
    P_{ij} = \frac{|x_i - x_j|^{-\beta}}{\sum_j{|x_i - x_j|^{-\beta}}},
\end{equation}

where $\beta$ is the homophily factor which quantifies the tendency for agents with similar opinions to interact with each other: $\beta = 0$ refers to the absence of interaction preference, and $\beta > 0$ implies that the agents with similar opinions are more likely to interact with one another. Evidently, Eq. \ref{homophily.eq} is 
modeled as a power-law decay of connection probabilities with only a small chance for agents with opposite opinions to interact. Since most of the interactions tend to occur between agents with similar opinions, this can lead to the formation of echo chambers.

\begin{figure}[ht]
    \label{pol_def.fig}
    \includegraphics[width=0.4\textwidth]{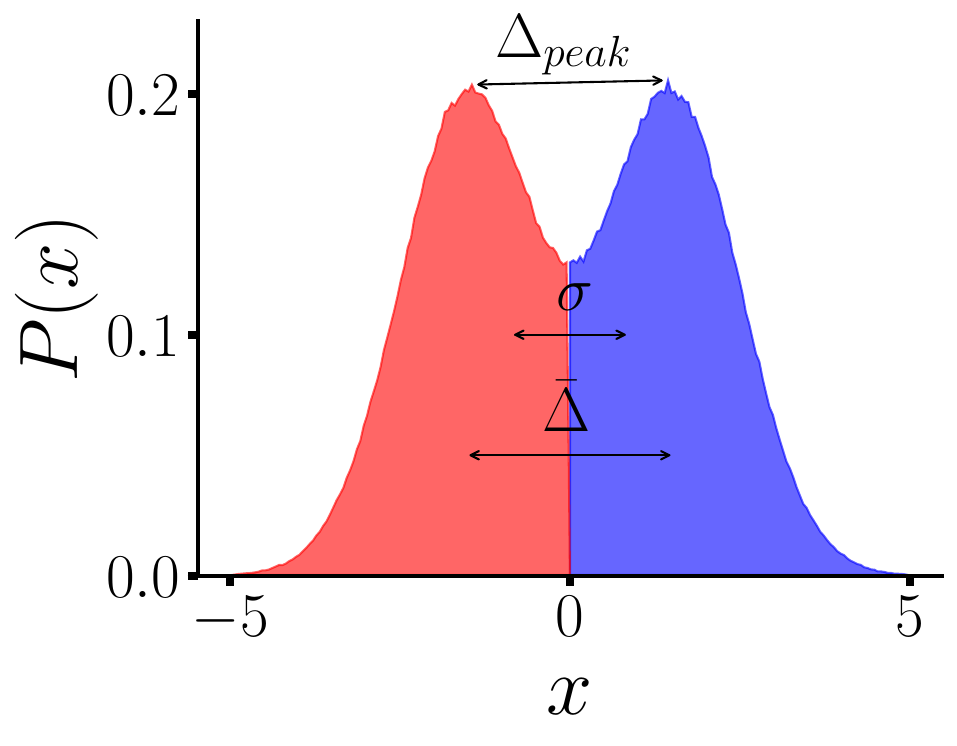}
    \caption{A schematic to illustrate three measures of polarization. \textbf{$\bar \Delta$} is the distance between mean positive and negative opinions.
    \textbf{$\Delta_{peak}$} denotes the distance between two peaks in the opinion distribution, and \textbf{$\sigma$} denotes the standard deviation of the opinion distribution.}
    \label{fig:pol_def}
\end{figure}

\begin{figure*}[!htbp]
    \includegraphics[width=\textwidth]{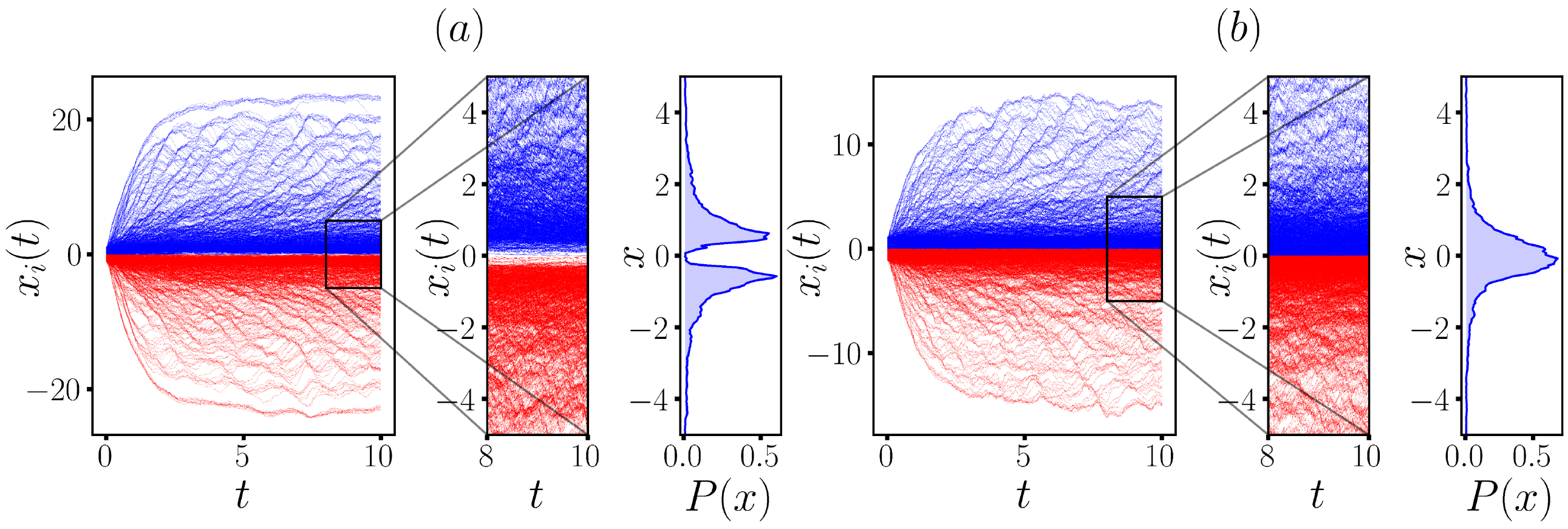}
    \caption{Emergent polarization (and depolarization) states in the presence (and absence) of the nudge factor. The simulations are performed with parameters set to be in the polarization regime. (a) The agents are not nudged. Hence the polarized state emerges. A magnification of the region around $x=0$ reveals the absence of trajectories there, and the corresponding distribution shows a bimodal distribution. (b) Network nudge is introduced with probability $p=0.01$, and we find a significant depolarization. Opinion trajectories tend to crowd around $x=0$, and a unimodal distribution emerges.}
    \label{fig:trajectory}
\end{figure*}
The interaction dynamics in the model is enforced by the activity-driven temporal network that is fully encoded 
by the parameters $(\varepsilon, \gamma, m, \beta, r)$, together with the parameters that characterises the issue, $(K, \alpha)$. Asymptotically, this model features three distinct states in the distribution of opinions. If the Social interaction $K$ is sufficiently small, then the opinion of every agent decays to zero, and this state is known as the neutral consensus state. However, if social interaction $K$ is large but the homophily factor $\beta$ is small, then due to statistical fluctuations, all the opinions either become positive or negative. This is the state of radicalization. And the most interesting case emerges when social interaction $K$ and homophily factor $\beta$ are large enough. In this case, the opinion distribution shows a bimodal polarized state.

\section{Random nudges and polarization}
\label{sec:random_nudge}
Echo chambers are increasingly becoming more apparent in online social media platforms. A generic tendency to interact with people who hold similar opinions as ours can lead to echo chambers, and this effect is, in turn, amplified by the recommendation engines on social media platforms. These algorithmically driven engines recommend similar connections or content in order to keep the users of those platforms engaged.

These two features are modeled by the interaction probability, controlled by the homophily factor $\beta$. Large values of $\beta$ represent how closed the echo chambers are. To disrupt the formation of echo chambers even while keeping the platforms as engaging as possible and without violating the users' privacy, we adopt the following intervention in the opinion dynamics model:
With probability $p<1$, the active agents will interact uniformly with any other agents, and with probability $(1 - p)$, the active agents will interact with others according to the homophily probability given in Eq. \ref{homophily.eq}. We call $p$ the random nudge probability. As $p$ does not depend on the opinions of the agents, the intervention is noninvasive (the recommendation engine need not interpret the opinion of the agents). For small enough values of $p$, it is hoped that the platform is still engaging while maintaining enough diversity to ensure there is no echo chamber. With this intervention, we propose a modified interaction probability as 
\begin{equation}
    \label{intervention.eq}
    \widetilde P_{ij} = p \times \frac{1}{N - 1} + (1 - p) \times P_{ij}.
\end{equation}
This is used in the rest of the results shown in this paper.

{\it \textbf{Quantifying Polarization}}: Before we delve into the details of the results, we discuss the three quantities employed to measure the degree of polarization based on the opinion distribution $P(x)$. They are defined as: ({\it a}) Polarization is measured through $\bar \Delta$, defined as the distance between the average of positive opinions and the average of negative opinions. {\it b}) When opinion distribution exhibits a bimodal character, the distance between the two peaks, denoted by $\Delta_{peak}$, can also be used as a measure of polarization \cite{depolarization-of-echo-chambers-by-random-dynamical-nudge}.
({\it c}) A gross measure of polarization could also be the standard deviation $\sigma$ of the entire opinion distribution \cite{link-recommendation-algorithms-and-dynamics-of-polarization-in-social-networks}. Fig. \ref{fig:pol_def} illustrates the schematics of all three measures of polarization. It must be noted that if polarization decreases due to the intervention proposed in Eq. \ref{intervention.eq}, ideally, all these three quantifiers must decrease.

\begin{figure*}[ht]
    \includegraphics[width=\textwidth]{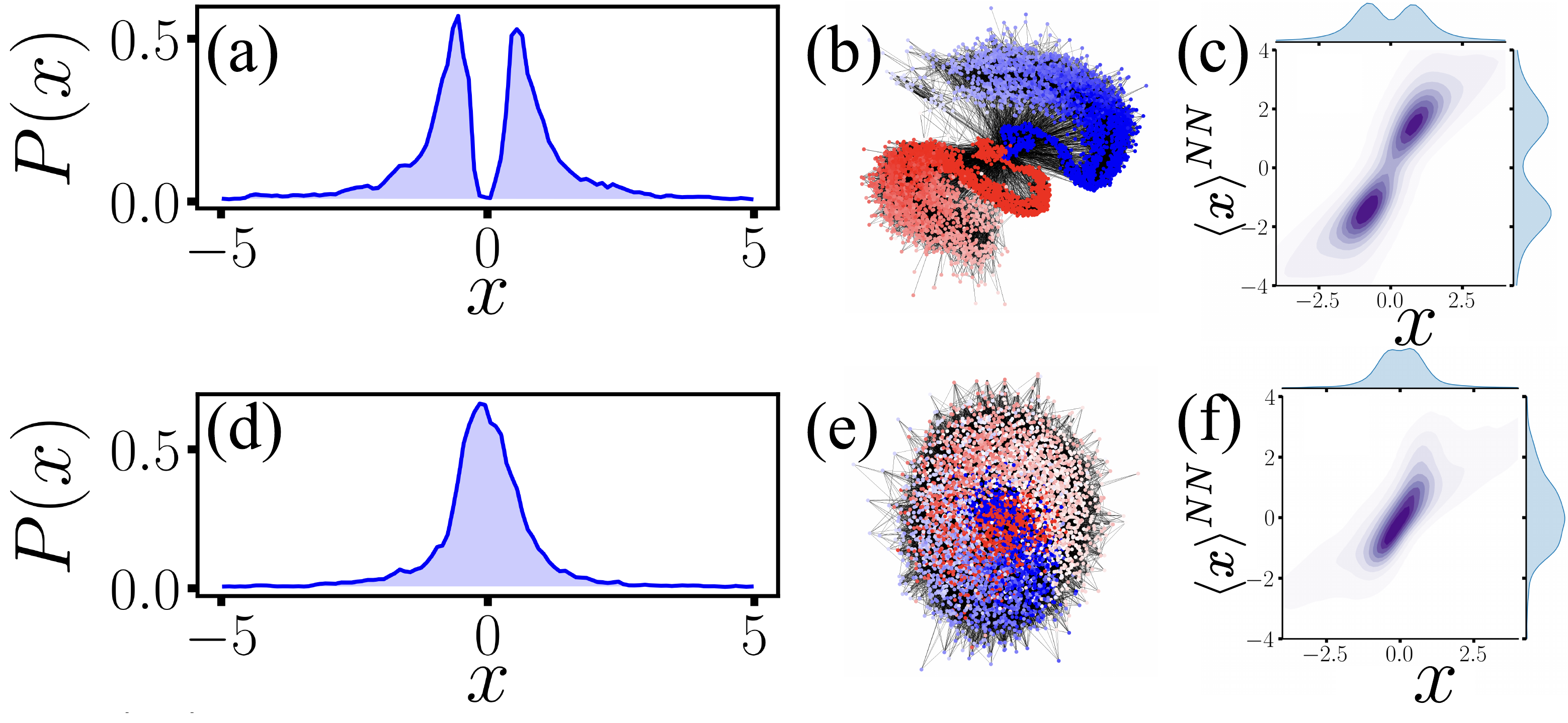}
    \caption{Effect of the nudge on the opinion distribution, the structure of social interactions networks, and the signature of echo chambers. The  networks are averaged over the last $100$ time steps of simulation and are drawn using the {\tt draw} function in {\tt networkx} \cite{networkx}. Nodes with 
    blue color correspond to agents with positive opinions, and red corresponds to agents with negative opinions. The saturation of the color is mapped to the conviction of the agents; high saturation corresponds to a high level of conviction, and vice-versa. The opinion of an agent $x$ and the mean opinion of its nearest neighbors $\langle x \rangle^{NN}$ is averaged over $20$ realizations to generate the contoured heatmap to indicate the presence of echo chambers (see Eq. \ref{eq.xnn}). And the marginal distributions are shown in the corresponding axes. (a) For $p= 0$, {\it i.e.}, without a nudge, the distribution is polarized, and the network has two distinct clusters (b), one formed by the agents with positive opinions and the other by the agents with negative opinions. (c) The presence of two lobes in the heatmap indicates the echo chamber effect. (d) For $p = 0.01$, we observe a unimodal distribution of opinion, and the social interactions network is now well mixed (e). A depolarization state is reached. (f) A single lobe in the heatmap confirms the weakening of the echo chamber effect.
    }

    \label{fig:network}
\end{figure*}

\section{Results}
\label{sec:results}
With the intervention strategy introduced in Sec. \ref{sec:random_nudge}, we find that with sufficiently small random nudge probability $p$, we obtain significant depolarization in the opinion distributions characterized by a unimodal distribution along with the decay of all three measures of polarization. To see the effects of nudge, we perform numerical simulations of the basic model in Eq. \ref{main.eq} using the interaction probability given in Eq. \ref{homophily.eq} and the intervention model in Eq. \ref{intervention.eq}. The simulations are performed with $N=10000$ agents for $1000$ time steps with $dt=0.01$. At initial time $x_i$ is uniformly chosen from a small interval, {\it i.e.}, $x_i \in [-1,1]$ for $i=1,2 \dots N$. The model parameters are chosen to be $\alpha=3$, $\beta=3$, $K=3$, $m=10$, $\gamma=2.1$, $\varepsilon=0.01$ and $r=0.5$ for all the simulations unless mentioned otherwise. The parameters chosen for the simulations lead to a polarized state in the original model without intervention.

In Fig. \ref{fig:trajectory}, we show the contrast between the trajectories of individual opinions and the opinion distribution with and without the application of a nudge. In the absence of nudge ($p=0$), the simulation results in Fig. \ref{fig:trajectory}(a) show fewer trajectories with opinions $x_i \approx 0$. This leads to a bimodal distribution of opinions characteristic of a polarized state. In contrast, in Fig \ref{fig:trajectory}(b), a small nudge with a probability of $p = 0.01$ is applied, and we find significantly more trajectories with moderate opinions. This, effectively, is seen to lead to an absence of polarization, characterized by a unimodal distribution. The magnifications of the region around $x_i=0$ and its distribution (shown in Fig. \ref{fig:trajectory}) reveal a clear distinction between these two scenarios.

To examine the effect of network nudge, we analyze the underlying time-averaged structures of the temporal interactions network. Without nudge, the interaction network has two distinct clusters; most of the connections are among positive opinionated agents or negative opinionated agents. There exist very few connections between these two groups other than for the agents with extreme opinions.
This is expected since the agents with extreme opinions are also those who tend to be more active on social networks fora; hence on average, they form more connections. This enables them to be relatively more connected to the agents with opposing opinions. These results are visually depicted in Fig. \ref{fig:network} as two snapshots of evolving network diagrams. If $p=0$, no nudge is applied. In this case, as Fig \ref{fig:network}(b) shows, a polarized network, made up of two distinct blue and red-colored clusters, is formed. Blue color corresponds to nodes with $x > 0$, and red color to $x< 0$. The opinion distribution shown in Fig. \ref{fig:network}(a) confirms the existence of polarization. Panels (a), (b), (d), and (e) of Fig. \ref{fig:network} are generated from the simulations with $5000$ agents.


\begin{figure*}[ht]
    \includegraphics[width=\textwidth]{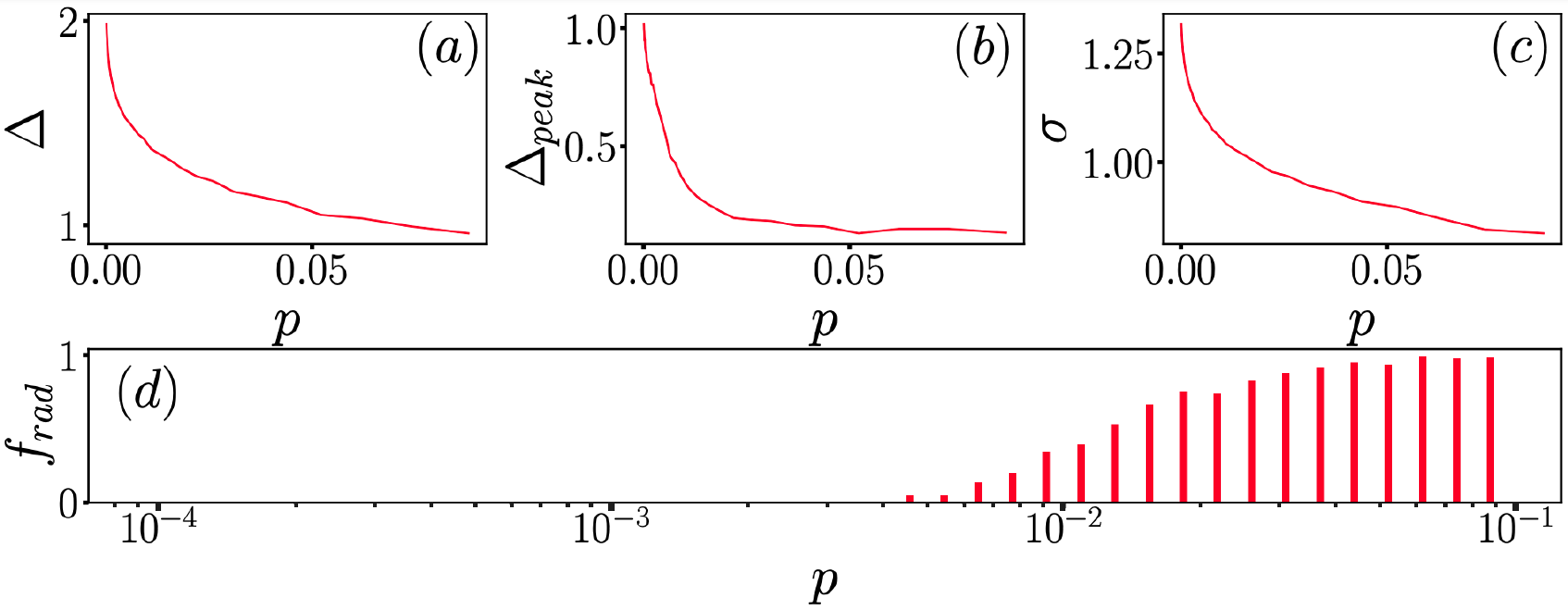}
    \caption{Three measures of polarization, (a) $\bar\Delta$, (b) $\Delta_{peak}$, and (c) $\sigma$,  as a function of nudge strength $p$. All three polarization parameters are averaged over the last 100 time steps and also averaged over 200 realizations. (d) shows the fraction of simulations that lead to radicalization for different nudge strengths.}
    \label{fig:pol_par}
\end{figure*}
However, when a nudge is applied, even for the case when the nudge probability is as small as $p = 0.01$, we find the network to be well mixed (large blue and red clusters have disappeared) (Fig. \ref{fig:network}(e)), and this leads to a depolarized state which is confirmed by the unimodality of the opinion distribution as shown in Fig. \ref{fig:network}(d).

\begin{figure*}[ht]
    \includegraphics[width=\textwidth]{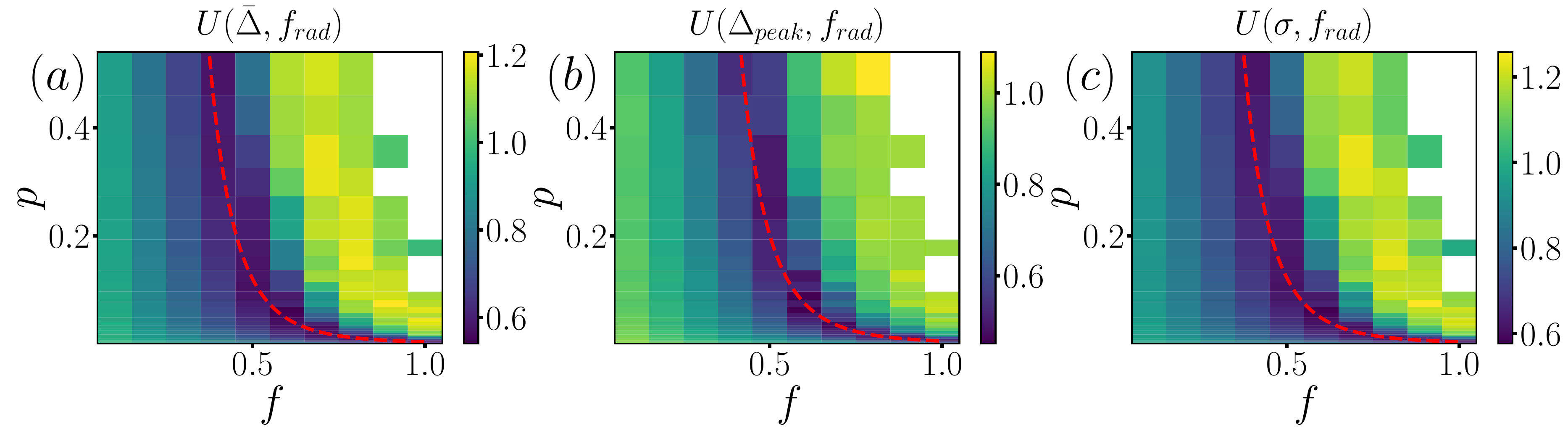}
    \caption{The heat map of the utility as a function of nudge strength and population fraction. Panel (a), (b), and (c) corresponds 
    to the corresponding utility of $\bar \Delta$, $\Delta_{peak}$, and $\sigma$, respectively. The red dashed curve, which is found to 
    follow the curve $p\cdot f^A = B$, ($A, B = $ constants), denotes the optimal values of population fraction and nudge strength.}
    \label{fig:optimization}
\end{figure*}
The term echo chamber describes a situation where the beliefs or opinions of people are reinforced by interactions among a closed group of people who hold similar opinions. In recent years, this has been widely discussed in the context of online communities \cite{echo-chambers-emotional-contagion-and-group-polarization-on-facebook, quantifying-echo-chamber-effects-in-information-spreading-over-political-communication, political-discourse-on-social-media-echo-chambers, The_echo_chamber_effect_on_social_media}. However, some studies appear to suggest that the effects of echo chambers are over-estimated \cite{the-echo-chamber-is-overstated}. To infer the presence of echo chamber-type effects, we calculate the average opinion of the nearest neighbors (NN) of each agent \cite{modeling-echo-chambers-and-polarizaiton-dynamics-in-social-networks, The_echo_chamber_effect_on_social_media}. This is denoted by 
\begin{equation}
\langle x\rangle^{NN} = k_i^{-1} \sum_j {a_{ij} x_j}, ~~~\mbox{and}~~~ k_i = \sum_j{a_{ij}},
\label{eq.xnn}
\end{equation}
where $a_{ij}$ is the temporally aggregated adjacency matrix. When a nudge is not applied ($p=0$), a colored contour plot of $x$ and $\langle x\rangle^{NN}$, in Fig. 
\ref{fig:network}(c) reveals two saturated spots corresponding to the two distinct echo chambers.
Now, when we apply a nudge with probability $p=0.01$, we can observe only one saturated spot indicating the existence of only one closed group (Fig.\ref{fig:network}(f)). All the agents are inside this closed group, and the echo chamber effect is largely diluted or non-existent. This is more clearly evident in the distributions plotted along the top and right-side axes of Fig. \ref{fig:network} (f). The panels (c) and (f) of Fig. \ref{fig:network} are generated from simulations with $1000$ agents.
\begin{figure}[ht]
\includegraphics[width=1\columnwidth]{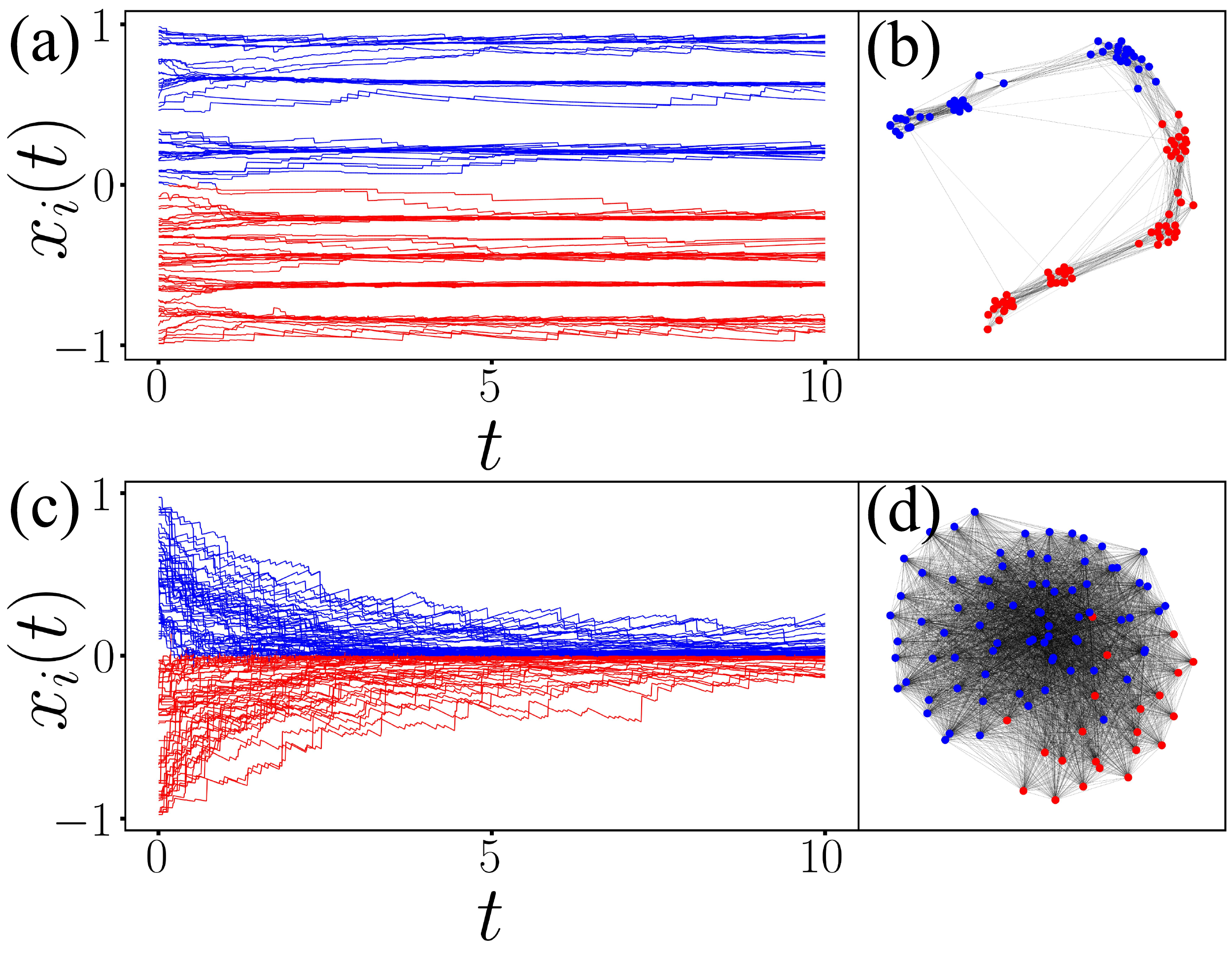}
    
    \caption{The effect of nudge in the opinion dynamics model, governed by equation \ref{new_model.eq}. Panels (a) and (c) show the trajectories of opinions in the absence and presence of network nudge, respectively. Panels (b) and (d) show the corresponding interaction network structure. Clearly, we see the presence of echo chambers in the absence of a network nudge, and the effect decreases when a slight nudge is applied.}
    \label{fig:new_model}
\end{figure}
\section{Optimizing the nudge: Polarization versus Radicalization} 
To obtain a global picture of how depolarization sets in as a function of nudge probability $p$, we plot the three measures of polarization as a function of $p$. All three measures, $\bar\Delta, \Delta_{peak}$ and $\sigma$, have been computed from the simulation results. The results shown represent an average over the last 100 time steps of simulation and averaged over 200 realizations. In Fig. \ref{fig:pol_par}, we observe that all three measures of polarization decrease as the strength of the nudge $p$ increases. In particular, $\bar \Delta$ and $\sigma$ are found to decrease as a stretched exponential function $\exp(-p^\gamma)$, and the stretching factor $\gamma$ is determined through regression to be approximately $0.3$. A recent work studying the  depolarization of echo chambers \cite{depolarization-of-echo-chambers-by-random-dynamical-nudge} considered adding an effective noise term dependent on a random sample of opinions to Eq. \ref{main.eq}. While this approach succeeds in making the opinion distribution unimodal, it increases the width of the distribution significantly, which as a consequence, corresponds to an increase in extreme opinions. In contrast, the framework of nudging the mechanism of forming social connections in online interactions works well in decreasing extreme opinions and also suggests direct algorithmic interventions for recommender systems.

One limitation of the intervention proposed in this work is that for large $p$, we observe that in a large fraction of the realizations, a radicalized state emerges. This can be seen in Fig. \ref{fig:pol_par}(d), which displays the fraction of realizations that lead to radicalization $f_{rad}$ as a function of $p$. It is clear that radicalization is either non-existent or a rarity for $p < 10^{-2}$, while radicalization becomes the norm for $p > 10^{-2}$. In many situations, radicalization is as much undesirable as polarization.
\par
To solve the issue of radicalization at a high value of nudge probability, rather than nudging all the people in the population, we nudged 
a fraction of the population. We define a simple linear utility function $U(\bar \Delta, f_{rad}) = \tilde{\bar \Delta} + f_{rad}$
Where $\tilde{\bar \Delta}$ is $\bar \Delta$, linearly scaled to be between $0$ and $1$, and $f_{rad}$ is the fraction of radicalized
simulations. The structure of the utility function is the same for the other two measures of polarization. In Fig. \ref{fig:optimization} 
we find the optimal values of population fraction and nudge probability, which follows the curve $p\cdot f^A = B$ ($A, B = $ constants).

\section{Robustness of the Framework}
To ensure the robustness of our intervention framework, we applied network nudge to another recent model \cite{modeling-explosive-opinion-depolarization-in-interdependent-topics} of opinion dynamics, which, together with homophily, exhibits the effect of echo chambers. The dynamics of the model is governed by the following $N$ coupled differential equations:
\begin{equation}
    \label{new_model.eq}
    \dot{x}_i(t)= |x_i|\sin{(x_i^0- x_i)} + K \left(\sum^{N}_{j=1} A_{ij} (t)  \sin{(x_i - x_j)}\right).
\end{equation}

In contrast to the original model \cite{modeling-explosive-opinion-depolarization-in-interdependent-topics}, the variable $x_i$ is chosen to be the opinions of the people on a single topic, and the temporal adjacency matrix is formed according to homophily probability \ref{homophily.eq}. $x_i^0$ is the initial opinion of agent $i$, and all the other variables and parameters have the same meaning as in the previous model \ref{main.eq}. In fig. \ref{fig:new_model}, we show that When the social interaction and the homophily factor are high enough ($K = 4$, $\beta = 4$), many echo chambers are formed, which is clear from the trajectories of the opinion as well as from the multiple communities seen in the aggregated network (Fig \ref{fig:new_model} (a, b)). But when we introduce a slight nudge ($p = 0.002$), The effect of echo chambers is reduced drastically. The opinion trajectories seem to converge to a moderate value, and the interaction network is well-connected without any obvious segregated communities (Fig \ref{fig:new_model} (c, d)).

\section{Discussion}

The widespread use of the internet, and consequently, social media platforms, have drastically altered the way humans consume and interact with information. Polarization and the formation of echo chambers have been shown to negatively impact constructive discussions and debates -- two fundamental pillars of a healthy democracy. Building on the recent advances in the modeling of opinion dynamics in social networks, in this work, we study the possibility of depolarizing a population using a stochastic nudge. 

Our results suggest that a small number of randomized interactions, which are other dominated by homophily-driven mechanisms, can lead to a significant reduction in polarization. This reduction was quantitatively captured by three different measures of polarization. While we show that minimal nudges can burst echo chambers and lead to socially desirable distributions of opinions, increasing the strength of this nudge can result in radicalization. Given this sensitivity on the nudge strength, we show that a possible resolution is obtained if, instead of nudging each agent, only a fraction $f$ of the agents are nudged. We highlight that this interplay of the nudge strength $p$ and the fraction $f$ of nudged individuals leads to an interesting optimization problem. This optimization can help inform the fraction of individuals to be nudged for a fixed nudge strength for optimal depolarization.

We believe that the strongest case for the application of such randomized nudges can be made to recommendation systems. While ubiquitous, recommender algorithms are optimized for increasing engagement \cite{recommender-systems-and-their-ethical-challenges}, which we now know can come at the cost of creating echo chambers \cite{echo-chambers-in-collaborative-filtering-based-recommendation-systems}, increase in the representation of extreme ideologies \cite{recommender-systems-and-the-amplification-of-extremist-content}, and even the tampering of users' preferences \cite{user-tampering}. In such settings, the randomized nudges can be potentially operationalized as the poisoning of a viewer's watch history with a limited amount of random content, uncorrelated with the viewer's preferences \cite{youtube-audit}. While there are several ethical and legal considerations that must be accounted for before implementing any such interventions, it certainly opens up several interesting avenues for future research to build on.

\acknowledgements{R.P. and A.K. gratefully acknowledge the Prime Minister’s Research Fellowship of the Government of India for financial support.  M.S.S. acknowledges the support of a MATRICS Grant from SERB, Government of India. The authors acknowledge the National Supercomputing Mission for the use of PARAM Brahma at IISER Pune.}

\end{document}